# Graphene-based optical sensors for the prevention of SARS-CoV-2 viral dissemination

Giuseppina Simone

The Ministry of Education Key Laboratory of Micro/Nano Systems for Aerospace, School of Mechanical Engineering, Northwestern Polytechnical University, 127 West Youyi Road, Xi'an Shaanxi, 710072, People's Republic of China.

**Abstract.**

What level of scientific knowledge has been gained against COVID-19 during the months of the pandemic? Are current technologies able to keep up with the advances in virology research?

Historically, disease outbreaks due to pathogenic microorganisms have killed many more people than wars have, and the ability that viral genes have displayed to rapidly change and adapt has forced new studies. Early detection is most important for timely management of any biological attack, whether natural or intentional, and rapid detection systems are essential to counter the effects of viral disease. Optical sensors have several advantages in this field, such as simple and label-free protocols.

In this letter, the role of layer materials, such as graphene, is presented. Graphene has unique and excellent optical properties for plasmonics and Raman scattering.

Finally, a method of Covid-19 detection on graphene sensors is proposed via a plasmon-molecular orbital interaction that provides a unique signature of the S-protein of the virus.

Traditionally, the detection of pathogens has played an important role in the field of research and safeguarding countries. Although microorganisms are essential for human life and the environment, historically, disease outbreaks due to pathogenic microorganisms have killed many more people than wars have (Hoyert DL, Kochanek KD, Murphy SL. 1999). A significant fear has arisen from the threat of biological warfare, defined as a planned and deliberate use of pathogenic strains of microorganisms such as bacteria, viruses, or their toxins, in order to spread life-threatening diseases on a mass scale and devastate the population of an area. Notably, uncommon diseases are used in biological warfare and, thus, populations are more susceptible to rapid infection.

At the end of 2019, cases of pneumonia of unknown etiology were detected in Wuhan, China, which led to an outbreak that the WHO declared a Public Health Emergency of International Concern in March 2020. Within a few months, 188 countries faced the COVID-19 disease, which has led to more than 21 million deaths with an overall mortality of approximately 3.6%, according to updates from the middle of August (the Center for Systems Science and Engineering (CSSE) at Johns) [COVID-19 Dashboard by the Center for Systems Science and Engineering (CSSE) at Johns Hopkins University (JHU)].

The COVID-19 disease is caused by Coronaviruses (CoVs) (**Figure 1a**), which are 100 nm viruses protected by twisting strands of RNA that grab human cells via the S-protein (Mark Fischetti 2020). In 2002 a coronavirus SARS-CoV-1 virus that caused severe acute respiratory syndrome killed approximately 1000 people worldwide; additionally, since 2012, a different strain producing Middle East respiratory syndrome (MERS) has taken more than 800 lives. SARS-CoV-2 has emerged as a new species able to infect humans and is currently causing the COVID-19 pandemic.

In the meantime, governments around the world have planned pandemic defense strategies including methods, protocols, and procedures aimed at establishing and executing defensive measures (detection, protection, decontamination, and medical management). The scientific community has worked to develop early detection methods to counter the disease spread and have generated a significant amount of knowledge in a surprisingly short time.

Early detection is most important for timely management of any biological attack, whether natural or intentional, and rapid detection systems are essential for countering viral disease. Standard methods of viral detection based on polymerase chain reaction, as well as electrochemical assays, require trained staff and expensive reagents and equipment to ensure reliable and reproducible sensitivity.

Therefore, more easily approachable methods are needed in order to achieve rapid and efficiently detection of the virus.

Conventional techniques for the detection of pathogens are demanded to enable an easy computation of signals.

Molecular detection systems hold huge potential as a tool for discovering the basic functions of biological molecules and the mechanisms of various diseases, however, the required performance levels of the detection systems are severe. For example, high-sensitivity systems are required since low concentrations of samples (e.g. 100 particles/L of *B. anthracis* or 10 particles/L of *F. tularensis*) are sufficient to cause human infection (Burrows and Renner 1999). At the same time, the specificity of the sensor must allow discriminating from bacterial contamination, as well as other biological and non-biological components, in order to reduce false-positive rates (Wilson et al. 2002). As can be seen from the receptors enumerated in **Table 1**, countless methodologies have been exploited and significant progress has been reported in this field and the operating steps involved in developing a pathogenic microbe-detecting sensor.

The design of the substrate architecture is fundamental for rapid and sensitive analysis and, thus, continues to motivate research.

A literature search of published works using the keywords 'SARS-CoV-2, optical sensors' in a time range spanning from December 2019 to August 2020 showed that 1640 articles had reported progress in this context. Next, a more specific search based on the keywords 'SARS-CoV-2, optical sensors, graphene' highlighted 130 studies that focused on the implementation of optical sensors using layered materials, such as graphene and its derivatives (Palmieri et al. 2019; Chauhan et al. 2017). Indeed, graphene components have offered solutions for fighting the current pandemic such as in effective antiviral disinfectants, surface coatings, impermeable masks, and sensors.

Graphene is an excellent candidate for biosensors due to its unique mechanical, physical, electrical, and optical properties. It has exceptional and unique properties (Song et al. 2015). It is a crystalline allotrope of sp2-hybridized carbon atoms, covalently linked and forming a honeycomb lattice structure. The large surface area of graphene reaches 2,630 m² g⁻¹, while the electron mobility is as high as 10,000–50,000 cm² V⁻¹ s⁻¹ at room temperature and the intrinsic mobility limit exceeds 200,000 cm² V⁻¹ s⁻¹ and is less influenced by temperature changes. The electrical conductivity of graphene could be up to $10^8$ mS cm⁻¹. The high surface area of graphene can be exploited for building multivalent substrates decorated by specific receptors (Seo et al. 2020) and the native hydrophobic (Munz et al. 2015) graphene can be converted to hydrophilic through oxidation (Hummers and Offeman 1958). This last mechanism has been pointed to as a key property of graphene/graphene oxide in facilitating control of the interaction with glycans used by viruses to hide their killing functions (Li et al. 2016).

Most of the optical properties of naïve graphene rely upon the *quasi*particles i.e. the phonons and are independent of the composition (**Figure 1b**). The phonons are described by a linear dispersion relation, which fixes a chiral-symmetry and the direction of pseudospin (e.g. parallel or antiparallel to the directions of motion of electrons and holes).

The exceptional phonon transport properties allow observation of the plasmonic fields and the coupling electromagnetic radiation in the range of the terahertz frequency, meaning that the photonic phenomena can be observed at a range of significantly low frequencies (Jung et al. 2010; Kirill V. Voronin et al. 2020).

Narrowing our attention to Raman scattering behavior, studies have shown that the enhancement because of graphene occurs if the highest occupied molecular orbital (HOMO) and the lowest unoccupied molecular orbital (LUMO) levels of that given molecule sit in a suitable energy range with respect to the Fermi energy $E_F$ of graphene (Feng et al. 2016; Huang et al. 2015). The variation in energy levels between the HOMOs and LUMOs strengthen the Raman scattering properties of graphene.

Doping the graphene alters its plain optical properties, with a significant modulation of optical transmission in the visible spectrum; and again, doping the graphene structure with other atoms, such as nitrogen or silicon, fosters a strong enhancement of the Raman behavior, which relies upon the charge transfer enhancement of which the mechanism illustrated in and it depends upon the level of doping as well as on the photons' energy.

The enhancement derived from doping graphene is significant, making clear that graphene sensors for Raman spectroscopy are capable of detecting trace amounts of molecules and viruses (concentration as low as $10^{-11}$ M), and may have a huge impact on hastening viral detection during this pandemic (Feng et al. 2013).

The optical phenomenon involved relies upon the non-radiative dipole-dipole interaction, that is the fluorescence resonance energy transfer between graphene and fluorescent substances; physically, the process is based on the non-radiation energy transfer between the donor and the acceptor separated by an opportune gap, in which the donor's emission spectrum overlaps the acceptor's absorption spectrum.

A question that needs addressing relates to the specific molecular interaction between the S-protein and the resonant Raman graphene orbitals. How are they affected by electron mobility? If this interaction between graphene and SARS-CoV-2 generates a specific Raman spectrum (Palmieri et al. 2019), the protocol becomes faster and easier (Fan et al. 2013; Mycroft-West et al. 2020).

Plasmonic detection of proteins has been successfully demonstrated. Recently, we have proposed the mechanism described in Figure 1c to explain the interaction between the hemoglobin protein and nanostructured Ag, which generates a strong, localized surface plasmon resonance signal (Zhang et al. 2020). The enhancement of the electromagnetic fields relies upon the excitation of propagating surface plasmon polaritons (SPPs), while the over-position of the plasmonic clouds intensifies the sensing capability in proximity to the metallic wires. The scheme in Figure 1c evidences that the molecular excitons from the proteins within the plasmon field formed strong coupling with the SPPs. The coupling of the exciton and the plasmon resonance yields the new chemical configuration, characterized by the new electronic states, which are derived from the interactions between the metal and the adsorbed molecules, having a unique HOMO/LUMO configuration that can provide uniqueness to the molecular readout.

An identical mechanism can be proposed for describing the interaction between the proteins and layered materials, like the graphene and its derivatives in contact with noble metals. A scheme of the mechanism for the graphene is reported in Figure 1d, which is based on the hypothesis that the interparticle distance is lower than the nanostructure's characteristic size and, consequently, a plasmonically induced electric field localized in the proximity of the metallic nanostructures is generated by the superposition of the plasmonic clouds. However, the characteristic structure of the graphene or the graphene oxide offers the electronic hole where the photon-excited conduction electrons jump and the oscillator–hole coupling is generated whereby the SPPs are excited at the metal–GO interface. The combination of the exciton and the plasmon resonance yields a new chemical configuration determined by the new electronic states derived from the interaction between the metal and the $\pi-\pi^*$ of GO.

The mechanisms reported above refer to the material's plasmonic behavior. However, under the hypothesis of a strong coupling between the SPPs and the electromagnetic field, they are still valid for describing the Raman scattering, as well as the surface enhanced Raman scattering, namely SERS, based on the enhancement given by the localized surface plasmon resonance and most importantly allow the molecular detection.

For example, the beam laser travelling through the prism facilitates matching of the photon and the surface plasmon polaritons SPP wavevectors by tunneling the photon in the total internal reflection geometry (Simone and Ruijter 2020). There, the coupling is strong and makes it possible the readout of the SERS by exploiting the molecular interaction with the plasmonic clouds, as presented in Figure 1d.

In conclusion, the validity of such a mechanism permits the selective detection of all molecules, as each of them is characterized by an individual and unique molecular configuration. Therefore, even homologue proteins, which include similar sequences of amino acids, can be discriminated, and the extent of discrimination depends upon the sequence similarity.

7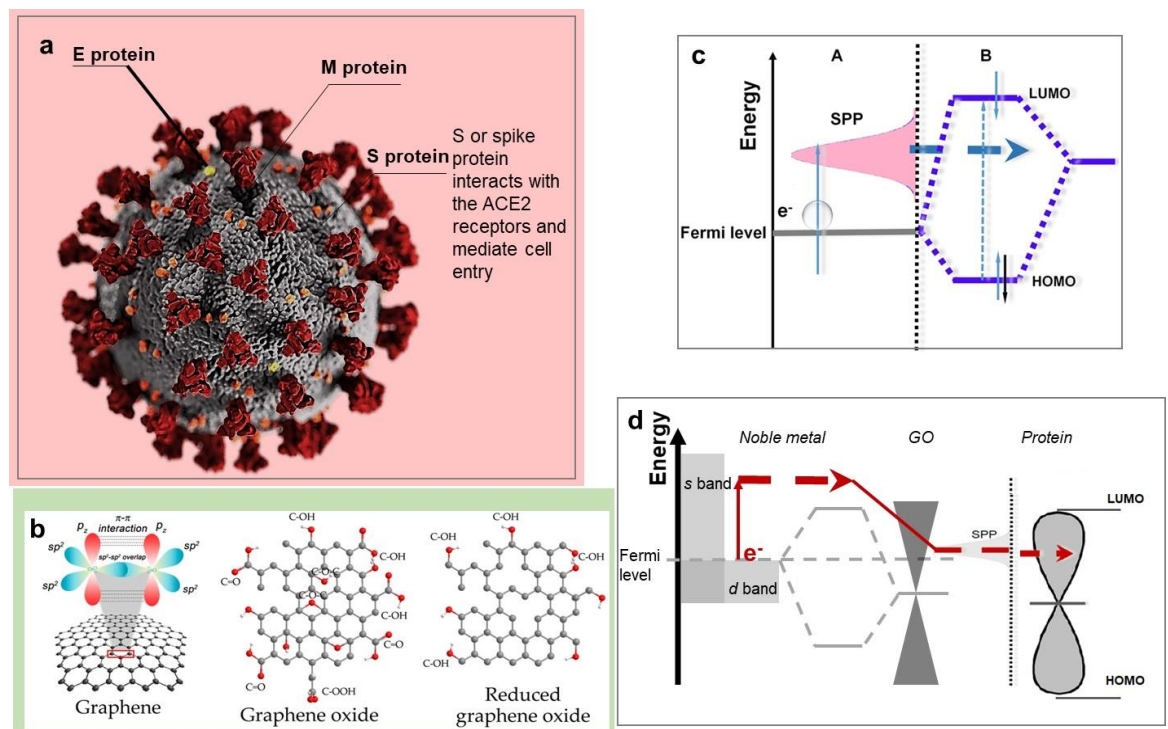

**Fig. 1:** (a) COVID-19 image at the microscopic level. Receptors, such as the S-protein are highlighted. (b) Molecular structure of graphene and its derivatives (Suvarnaphaet and Pechprasarn 2017). (c) Plasmonic interaction of metal and a protein (Zhang et al. 2020). (d) A proposed model regarding the behavior of graphene. The Fermi level of graphene can shift, and alignment of this shift with the lowest unoccupied molecular orbital of a protein enhances the charge transfer and amplifies the molecule's vibrational Raman modes.

**Table 1:** Operating steps in developing a pathogenic microbe-detecting sensor

| Receptor | Substrate | Signal amplification method | Sensing |
|---|---|---|---|
| Nucleic acid | Glass | Target recycling reaction | Colorimetric |
| Functional nucleic acid | Paper | DNase based reaction | Fluorescence |
| Antibody | Nano or layered materi- | Magnetosome amplification | SERS |
| Antimicrobial peptide | als | method | SPR |
| Carbohydrate | Polymer | | |
| Bacteriophage | Silicon | | |